\documentclass[aps,prx,reprint,superscriptaddress]{revtex4-2}

\usepackage{amsmath}
\usepackage{upgreek}
\usepackage{bm}
\usepackage{float}
\usepackage{graphicx}
\usepackage{tabularx}
\usepackage{hyperref}

\usepackage{subcaption}

\begin{document}

\title{High-Dimensional Neural Network Potentials for Magnetic Systems Using Spin-Dependent Atom-Centered Symmetry Functions}

\author{Marco Eckhoff}
\email{marco.eckhoff@chemie.uni-goettingen.de}
\affiliation{Universit\"at G\"ottingen, Institut f\"ur Physikalische Chemie, Theoretische Chemie, Tammannstra{\ss}e 6, 37077 G\"ottingen, Germany.}
\author{J\"org Behler}
\email{joerg.behler@uni-goettingen.de}
\affiliation{Universit\"at G\"ottingen, Institut f\"ur Physikalische Chemie, Theoretische Chemie, Tammannstra{\ss}e 6, 37077 G\"ottingen, Germany.}
\affiliation{Universit\"at G\"ottingen, International Center for Advanced Studies of Energy Conversion (ICASEC), Tammannstra{\ss}e 6, 37077 G\"ottingen, Germany.}

\date{\today}

\begin{abstract}
Machine learning potentials have emerged as a powerful tool to extend the time and length scales of first principles-quality simulations. Still, most machine learning potentials cannot distinguish different electronic spin orientations and thus are not applicable to materials in different magnetic states. Here, we propose spin-dependent atom-centered symmetry functions as a new type of descriptor taking the atomic spin degrees of freedom into account. When used as input for a high-dimensional neural network potential (HDNNP), accurate potential energy surfaces of multicomponent systems describing multiple magnetic states can be constructed. We demonstrate the performance of these magnetic HDNNPs for the case of manganese oxide, MnO. We show that the method predicts the magnetically distorted rhombohedral structure in excellent agreement with density functional theory and experiment. Its efficiency allows to determine the N\'{e}el temperature considering structural fluctuations, entropic effects, and defects. The method is general and is expected to be useful also for other types of systems like oligonuclear transition metal complexes.
\end{abstract}


\maketitle


In recent years, machine learning potentials (MLP), which allow to extend the time and length scales of first principles-quality atomistic simulations \cite{Behler2016, Bartok2017, Noe2020}, have become a rapidly growing field of research. More and more complex systems have been investigated driving new developments and extending the applicability of MLPs. Many of the current MLPs can be classified into four generations \cite{Ko2020a, Behler2021}: The first generation of MLPs proposed already in 1995 \cite{Blank1995} typically employs a single or a few feed-forward neural networks making them applicable to low-dimensional systems like small molecules in vacuum or diatomic molecules interacting with frozen surfaces of solids \cite{Handley2010, Behler2011a}. In 2007, second-generation MLPs have become available with the introduction of high-dimensional neural network potentials (HDNNP) \cite{Behler2007, Behler2014, Behler2015, Behler2017}. These MLPs are applicable to systems containing thousands of atoms by making use of the locality of a large part of the atomic interactions. For this purpose, the potential energy is calculated as a sum of local environment-dependent atomic energies defined by a cutoff radius. Most modern MLPs belong to this second generation, for example, HDNNPs \cite{Behler2007, Smith2017}, Gaussian approximation potentials \cite{Bartok2010}, moment tensor potentials \cite{Shapeev2016}, and many others \cite{Balabin2011, Thompson2015, Drautz2019}. The third generation of MLPs includes long-range interactions, mainly electrostatics but also dispersion, beyond the cutoff radius. The electrostatic interactions can be based on element-specific fixed charges \cite{Bartok2010, Deng2019} or local environment-dependent atomic charges expressed by machine learning \cite{Artrith2011, Morawietz2012, Yao2018}. Also message passing networks with explicit electrostatics have been proposed \cite{Unke2019}. Finally, fourth-generation MLPs take non-local or even global dependencies in the electronic structure into account, and consequently the atomic charges can adapt to non-local charge transfer and even different global charge states. The first method of this generation has been the charge equilibration neural network technique (CENT) \cite{Ghasemi2015}, and further methods like Becke population neural networks (BpopNN) \cite{Xie2020} and fourth-generation (4G) HDNNPs \cite{Ko2020} have been introduced recently. 

In spite of these methodical advances, which extended the complexity of the systems that now can be studied and the physical phenomena that can be included, a remaining limitation of MLPs is the inability to take different spin orientations and thus magnetic interactions into account. The reason for this limitation is that typically MLPs are trained to represent the potential energy surface of one electronic state as a function of structural descriptors -- most often but not necessarily the ground state of a system. With the exception of different global charge states in fourth-generation potentials, describing multiple electronic states usually requires to train separate MLPs for each state \cite{Behler2008, Dral2018, Chen2018, Hu2018, Wang2019} or to use one MLP yielding a vector of output energies corresponding to the different excited states \cite{Williams2018, Westermayr2019, Westermayr2020, Westermayr2020a, Westermayr2020b}. The unique structure-energy relation of a given state is a central component of nowadays MLPs, and energy changes resulting from different spin orientations give rise to the problem of contradictory information in the training process. Only very recently, magnetic moment tensor potentials (mMTP) \cite{Novikov2020} have been proposed as a first example of a MLP which is able to describe magnetic systems containing a single element like iron. 

For studying magnetic materials, the description of one magnetic state only is not sufficient as the atomic magnetic moments fluctuate and change signs at finite temperatures. These atomic spin-flips often play a role already at ambient temperatures as the energy differences between different spin configurations are typically only in the order of a few meV\,atom$^{-1}$. For example, the Curie and N\'{e}el temperatures, at which a material loses its ferromagnetic or antiferromagnetic state, are typically below 1000 and 500\,K, respectively \cite{Sanvito2018}. Magnetic transitions often give rise to structural changes \cite{Greenwald1950}. However, current implementations of MLPs are not able to capture these effects because the magnetic interaction is not included explicitly. For example, if an antiferromagnetic ground state is used for training the MLP, this state will also be the basis of atomistic simulations at higher temperatures irrespective of the true magnetic ground state under these conditions.

To calculate the energy contribution of a magnetic configuration in an atomistic simulation, models like the Ising model \cite{Ising1925}, the Heisenberg model \cite{Heisenberg1928}, and the Hubbard model \cite{Hubbard1963} are widely used. However, these models are based on lattices, and structural and spin changes at finite temperatures cannot be considered simultaneously. If both contributions are important, the only generally applicable option is the use of ab initio molecular dynamics, which is based on the explicit calculation of the electronic structure in each step and thus is able to distinguish different magnetic states. This approach, however, is inevitably associated with high computational costs restricting simulations to small systems and short time scales.

Beyond the field of potential-energy surfaces, several machine learning approaches have been developed to address the properties of magnetic compounds, for example for the prediction of magnetic moments \cite{Sanvito2018} and ordering temperatures \cite{Sanvito2017, Nelson2019, Nguyen2019} from structural parameters. Moreover, machine learning methods are able to classify ferromagnetic and antiferromagnetic ground state materials \cite{Long2021} and to predict spin state splittings and metal-ligand bond distances in transition metal complexes \cite{Janet2017, Janet2018}. Recently, a high-dimensional neural network approach for the prediction of oxidation and spin states has been developed \cite{Eckhoff2020b}. 

In spite of these applications of machine learning to magnetic compounds, with the exception of mMTPs \cite{Novikov2020} to date there is no method enabling large-scale first principles-quality atomistic simulations of systems explicitly including magnetic interactions. The main reason for this lack of methods is the use of descriptors in current MLPs which exclusively depend on the atomic structure, like atom-centered symmetry functions (ACSF) \cite{Behler2011}, smooth overlap of atomic positions (SOAP) \cite{Bartok2013} and many others \cite{Langer2020}. Exceptions are BpopNNs \cite{Xie2020} and the recently introduced fourth generation of HDNNPs \cite{Ko2020} in which the atomic charges are included as additional information besides the structural descriptors to provide qualitative information about the electronic structure.

Describing the spin configuration in a form suitable as input for MLPs is very challenging, as not only the absolute spins but also the relative orientations of the atomic spins and distances of the respective atoms are vital. Hence, to predict the energy and forces simultaneously as a function of the geometric and spin configuration, suitable spin-dependent descriptors are needed. Here, we propose such a descriptor to construct MLPs simultaneously applicable to multiple magnetic states based on a new type of spin-dependent atom-centered symmetry function (sACSF), which adds the description of the magnetic configuration for the case of collinear spin polarization. The sACSFs produce atomic energies as a function of the local geometric and spin environment and thus formally represent a second-generation potential that can also be combined with third and fourth generation MLPs to include additional physics like long-range electrostatic interactions. In this work we will benchmark sACSFs employing second-generation HDNNPs \cite{Behler2007}, extending these potentials by an explicit dependence on the full spin configuration space to describe magnetic interactions.

We choose manganese oxide, MnO, to assess the quality of the resulting magnetic HDNNPs (mHDNNP) that can be constructed using sACSFs, because of the well characterized antiferromagnetic ground state configuration \cite{Shull1949, Shull1951}, the N\'{e}el temperature of $T_\mathrm{N}^\mathrm{exp}=116\,\mathrm{K}$ \cite{Bizette1938, Siegwarth1967}, and the rhombohedral distortion of the antiferromagnetic phase with lattice constant $a^\mathrm{exp}=4.430\,\mathrm{\AA}$ and lattice angle $\alpha^\mathrm{exp}=90.62\,^\circ$ at 8\,K \cite{Shaked1988} making this system a very interesting and challenging benchmark case. The magnetic unit cell is a $2\times2\times2$ supercell of the geometric unit cell and is built from (111) planes of ferromagnetically coupled Mn ions \cite{Shull1951}. These planes couple antiferromagnetically to the neighboring planes. Spin-polarized density functional theory (DFT) calculations employing the hybrid functional PBE0 \cite{Perdew1996a, Adamo1999} yield the correct magnetic ground state, called AFM-II, with the rhombohedral lattice parameters $a^\mathrm{PBE0}=4.40\,\mathrm{\AA}$ and $\alpha^\mathrm{PBE0}=90.88\,^\circ$ in good agreement with experiment \cite{Franchini2005}. The rhombohedral distortion is a consequence of the magnetic anisotropy, which arises from the magnetic dipole interactions \cite{Schroen2012}.

Several questions have not yet been conclusively answered for this system because the required large simulation cells and long time scales are inaccessible by DFT and the determination of coupling constants for lattice models gets very complicated as soon as defects have to be considered. Using a mHDNNP we are able to perform high-throughput studies of the equilibrium geometries of different magnetic configurations as well as to simulate the antiferromagnetic to paramagnetic phase transition to determine the N\'{e}el temperature. Moreover, the mHDNNP enables the inclusion of defects, as Mn vacancies, to simulate their role in real materials.

\section*{Results and Discussion}

\subsection*{Magnetic High-Dimensional Neural Network Potential}

The reference data set of the mHDNNP consists of 3101 $2\times2\times2$ bulk supercells of MnO and Mn$_{0.969}$O in various magnetic states and their corresponding HSE06 DFT energies and force components. The supercells include different displacements of the atomic positions from the ideal rock salt lattice as well as distortions of the lattice parameters. The construction of the reference data set is described in detail in the Supplementary Information \cite{SI}.

Training this data set with conventional ACSFs yields an energy root mean squared error (RMSE) of about 11\,meV\,atom$^{-1}$ in the best potential we were able to construct. This RMSE is an order of magnitude higher than the usual HDNNP accuracy of 1\,meV\,atom$^{-1}$ because the ACSFs can only provide geometrical information to assign the energy. Thus, HDNNPs based on ACSFs only predict an averaged potential energy regardless of the magnetic configurations as the best compromise while, for example, the HSE06 DFT functional yields an energy difference of 45.9\,meV\,atom$^{-1}$ between the AFM-II order and the ferromagnetic (FM) order for the ideal rock salt MnO structure using the experimental lattice constant $a_\mathrm{exp}$. 

\begin{figure}[htb]
\centering
\includegraphics[width=\columnwidth]{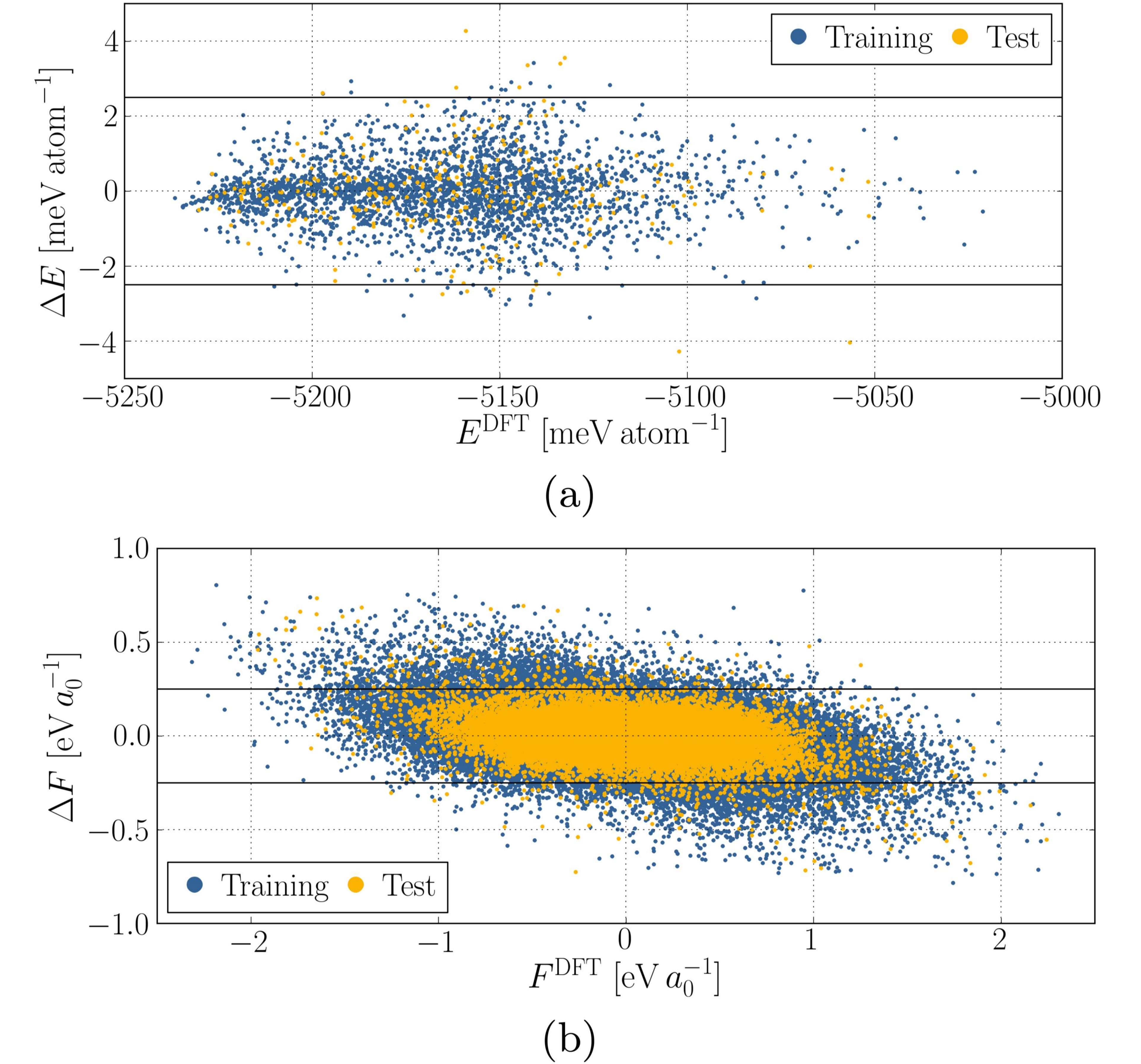}
\caption{(a) Energy errors $\Delta E$ as a function of the reference energy $E^\mathrm{DFT}$ and (b) force component errors $\Delta F$ as a function of the reference force components $F^\mathrm{DFT}$.}\label{fig:mHDNNP_qualtity}
\end{figure}

By including sACSFs to distinguish the magnetic configurations the energy RMSE is reduced by one order of magnitude to about 1\,meV\,atom$^{-1}$. Now, the energy difference between the AFM-II order and the FM order for the rock salt structure is resolved and is predicted to be 46.3\,meV\,atom$^{-1}$ in excellent agreement with the DFT reference data. Both results demonstrate the ability of the sACSFs to accurately describe the different magnetic configurations. The energy errors $\Delta E=E^\mathrm{mHDNNP}-E^\mathrm{DFT}$ and force component errors $\Delta F=F^\mathrm{mHDNNP}-F^\mathrm{DFT}$ are plotted as a function of the reference values in Figure \ref{fig:mHDNNP_qualtity} (a) and (b). The underlying number of Mn$_x$O reference structures and key performance indicators like RMSEs, maximum errors, and fractions of data points with high errors are compiled in Table \ref{tab:mHDNNP_quality} for the training and test data sets. It can be clearly seen that even when including various magnetic states the accuracy of the mHDNNP is in the typical region of state-of-the-art MLPs of 1\,meV\,atom$^{-1}$ and 0.1\,eV\,$a_0^{-1}$ for energies and forces with a test energy RMSE of 1.11\,meV\,atom$^{-1}$ and a test force components RMSE of 0.066\,eV\,$a_0^{-1}$. $a_0$ is the Bohr radius. In comparison, the RMSE of a recent magnetic moment tensor potential for defect-free body centered cubic iron restricted to a fixed lattice parameter is 2.0\,meV\,atom$^{-1}$ \cite{Novikov2020}, i.e., in the same order of magnitude. 

\begin{table}[htb]
\caption{Number of Mn$_x$O structures $N_\mathrm{struct}$ in the training and test set and the key performance indicators RMSE, maximum error, and fraction of data points with energy and force errors of the mHDNNP higher than 2.5\,meV\,atom$^{-1}$ and 0.25\,eV\,$a_0^{-1}$, respectively. The reference energy range is $-5236.6\,\mathrm{meV\,atom}^{-1}\leq E\leq-5021.2\,\mathrm{meV\,atom}^{-1}$ and the reference force components range is $|F|\leq2.31\,\mathrm{eV}\,a_0^{-1}$.}
\begin{center}
\begin{tabular}{lrrrr}
\hline
 &\,& Training set &\,& Test set\\
\hline
$N_\mathrm{struct}(\mathrm{MnO})$ && 1387 && 156\\
$N_\mathrm{struct}(\mathrm{Mn}_{0.969}\mathrm{O})$ && 1421 && 137\\
\hline
$\mathrm{RMSE}(E)$ && 0.86 && 1.11\\
$\Delta E_\mathrm{max}$ && 3.4 && 4.3\\
$\Delta E>2.5$ && 1.07\% && 4.44\%\\
\hline
$\mathrm{RMSE}(F)$ && 0.067 && 0.066 \\
$\Delta F_\mathrm{max}$ && 0.81 && 0.73\\
$\Delta F>0.25$ && 0.98\% && 0.89\%\\
\hline
\end{tabular}
\end{center}
\label{tab:mHDNNP_quality}
\end{table}

\begin{figure}[htb]
\centering
\includegraphics[width=\columnwidth]{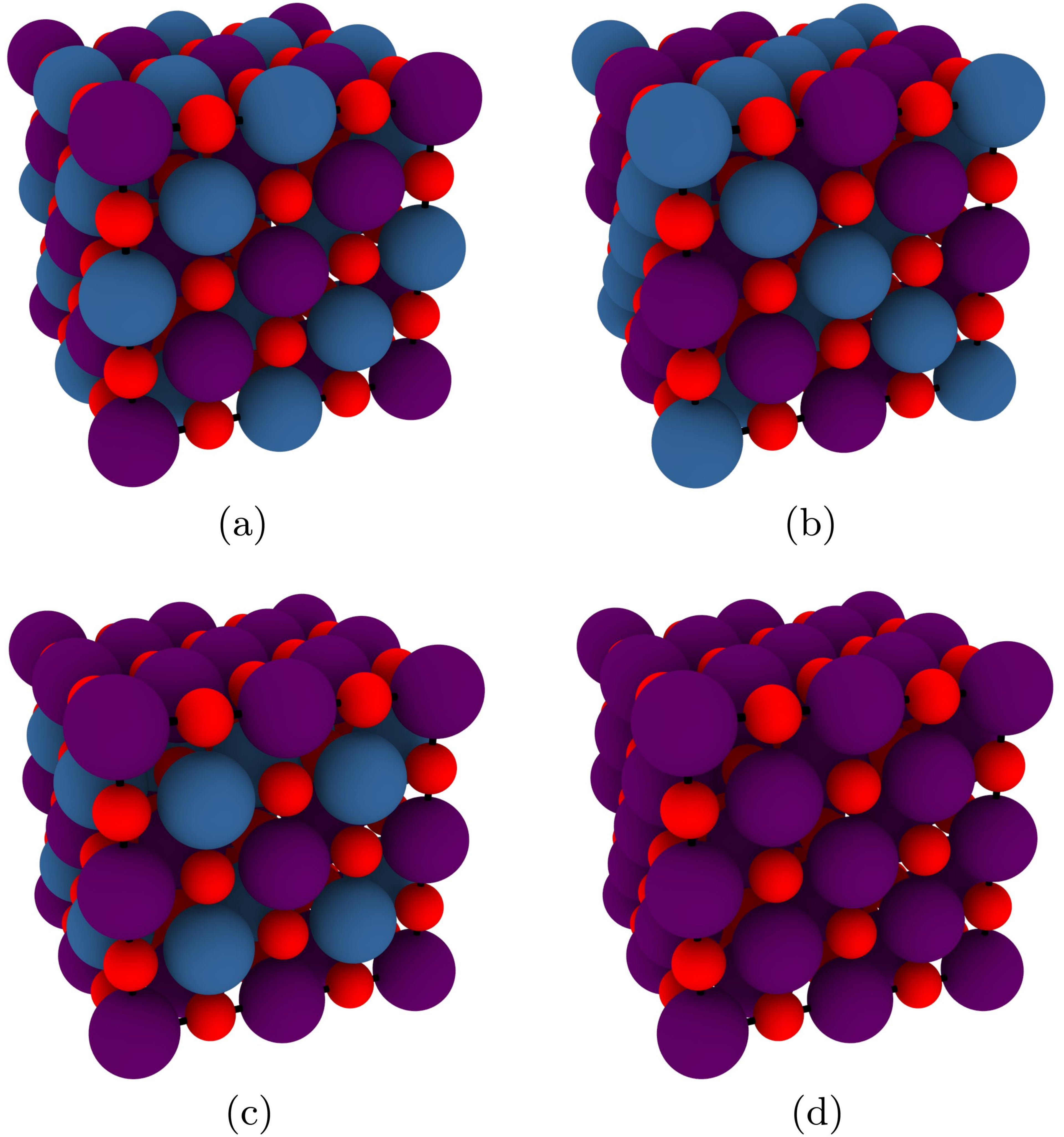}
\caption{(a) Rhombohedral MnO global minimum structure with AFM-II order, (b) magnetic order degenerate to the cubic AFM-II order, (c) AFM-I order, and (d) FM order of MnO. Small red balls represent O atoms, large violet ones Mn atoms with spin-up, and large blue ones Mn atoms with spin-down.}\label{fig:magnetic_order}
\end{figure}

\subsection*{Magnetic Configurations}

Besides the accurate description of the energies and forces the sACSFs enable to predict structural changes arising from the magnetic configuration. For example, for the rhombohedral MnO global minimum structure with AFM-II order (see Figure \ref{fig:magnetic_order} (a)) an unconstrained optimization yields the lattice parameters $a^\mathrm{mHDNNP}=4.433\,\mathrm{\AA}$ and $\alpha^\mathrm{mHDNNP}=90.77\,^\circ$ in very good agreement with the DFT results $a^\mathrm{DFT}=4.434\,\mathrm{\AA}$ and $\alpha^\mathrm{DFT}=90.89\,^\circ$. The corresponding experimental values are $a^\mathrm{exp}=4.430\,\mathrm{\AA}$ and $\alpha^\mathrm{exp}=90.62\,^\circ$ \cite{Shaked1988}. 

Also excited magnetic configurations are predicted in agreement with DFT. For example, the optimized lattice parameter of the resulting cubic lattice for the FM configuration (see Figure \ref{fig:magnetic_order} (d)) is $a^\mathrm{mHDNNP}_\text{FM}=4.461\,\mathrm{\AA}$ compared to the DFT value $a^\mathrm{DFT}_\text{FM}=4.462\,\mathrm{\AA}$. The energy difference to the global minimum is 45.8\,meV\,atom$^{-1}$ in excellent agreement with the DFT value 45.5\,meV\,atom$^{-1}$. The longer distances between ferromagnetically interacting Mn ions compared to antiferromagnetically interacting ones in MnO are emphasized by the optimized lattice parameters of the AFM-I configuration (see Figure \ref{fig:magnetic_order} (c)) with tetragonal lattice parameters $a^\mathrm{mHDNNP}_\text{AFM-I}=4.461\,\mathrm{\AA}$ and $c^\mathrm{mHDNNP}_\text{AFM-I}=4.414\,\mathrm{\AA}$ ($a^\mathrm{DFT}_\text{AFM-I}=4.459\,\mathrm{\AA}$ and $c^\mathrm{DFT}_\text{AFM-I}=4.420\,\mathrm{\AA}$), since the lattice is elongated in both directions of ferromagnetic interactions to the nearest neighbours. 

In conclusion, the lattices of FM (cubic), AFM-I (tetragonal), and AFM-II orders (rhombohedral) as well as other magnetic configurations can be different due to the magnetic interaction. This cannot be described by a lattice model such as the Heisenberg spin Hamiltonian. Further, the mHDNNP can provide information about the influence of defects in the magnetic order. For example, one spin-flip in the AFM-II configuration per $2\times2\times2$ supercell reduces the rhombohedral distortion by $0.09\,^\circ$ (DFT: 0.11\,$^\circ$). The corresponding energy increase is 1.5\,meV\,atom$^{-1}$ (DFT: 1.5\,meV\,atom$^{-1}$).

The efficiency of the mHDNNP in combination with a basin-hopping Monte Carlo scheme \cite{Wales1997} in which Monte Carlo (MC) spin-flips are employed instead of atomic displacements enables high throughput searches of the minima in spin configuration space which would be computationally too demanding employing DFT directly. The spin-flip basin hopping Monte Carlo (SFBHMC) simulations of $2\times2\times2$ MnO supercells as well as molecular dynamics (MD) simulations including MC spin-flips (MDMC) of $6\times6\times6$ MnO supercells confirmed the rhombohedral AFM-II magnetic order to be the global minimum in agreement with experiments \cite{Shull1951}. However, if the lattice is restricted to be cubic, two degenerate global minima exist: The AFM-II magnetic configuration shown in Figure \ref{fig:magnetic_order} (a) and another antiferromagnetic order shown in Figure \ref{fig:magnetic_order} (b). A recalculation of the structure using DFT confirms this observation. Radial distribution functions (see Supplementary Information \cite{SI}) show that the same ferro- and antiferromagnetic interactions are present in the two configurations. This second cubic global minimum configuration stays cubic in an unconstrained optimization. The rhombohedral distortion of the AFM-II order can therefore be identified as the origin of the energetic preference. The energy gain by the rhombohedral distortion is 1.9\,meV\,atom$^{-1}$ (DFT: 2.1\,meV\,atom$^{-1}$).

\subsection*{Magnetic Interactions}

For the ideal cubic structure a Heisenberg spin Hamiltonian can be constructed,
\begin{align}
H=-\dfrac{J_1k_\mathrm{B}}{2}\sum_i\sum_{n_i}\mathbf{S}_i\mathbf{S}_{n_i}-\dfrac{J_2k_\mathrm{B}}{2}\sum_i\sum_{m_i}\mathbf{S}_i\mathbf{S}_{m_i}\ , \label{eq:Heisenberg}
\end{align}
which includes the magnetic interactions of atom $i$ with its nearest neighbors $n_i$ and next nearest neighbors $m_i$. The strengths of the magnetic coupling between the vector spin operators $\mathbf{S}_i$ and $\mathbf{S}_{n_i}$ as well as $\mathbf{S}_i$ and $\mathbf{S}_{m_i}$ are given by the exchange coupling constants $J_1$ and $J_2$, respectively. $k_\mathrm{B}$ is the Boltzmann constant. The exchange coupling constants can be determined employing energetic differences among the FM, AFM-I, and AFM-II configurations, and Equation (\ref{eq:Heisenberg}) for these systems can be rearranged to yield
\begin{align}
J_1&=\dfrac{E_\text{AFM-I}-E_\text{FM}}{4S^2k_\mathrm{B}}\ ,\\
J_2&=\dfrac{4E_\text{AFM-II}-3E_\text{AFM-I}-E_\text{FM}}{12S^2k_\mathrm{B}}\ ,
\end{align}
with the spin $S=\tfrac{5}{2}$ of the high-spin Mn$^\mathrm{II}$ ions. Using mean field theory \cite{Ashcroft1976} the N\'{e}el temperature can then be calculated as
\begin{align}
T_\mathrm{N}=-2S(S+1)J_2\ .
\end{align}
Employing the HSE06 DFT energies using the experimental lattice parameter $a_\mathrm{exp}=4.430\,\mathrm{\AA}$ we obtain $J_1=-13.9\,\mathrm{K}$, $J_2=-14.5\,\mathrm{K}$, and $T_\mathrm{N}=255\,\mathrm{K}$. The mHDNNP results match these values almost perfectly with $J_1=-14.0\,\mathrm{K}$, $J_2=-14.6\,\mathrm{K}$, and $T_\mathrm{N}=256\,\mathrm{K}$. This agreement again underlines the performance of the sACSFs as well as of the mHDNNP method to describe the multiple potential energy surfaces of the MnO magnetic states. While the experimental N\'{e}el temperature of $T_\mathrm{N}^\mathrm{exp}=116\,\mathrm{K}$ \cite{Bizette1938, Siegwarth1967} is lower, this overestimation of the DFT value compared to experiment was also found in previous studies, for example, the PBE0 functional yields $T_\mathrm{N}=240\,\mathrm{K}$ \cite{Franchini2005} and the HSE03 functional $T_\mathrm{N}=230\,\mathrm{K}$ \cite{Schroen2010}.

\subsection*{N\'{e}el Temperature}

Mean field theory misses the influence of the specific magnetic configurations of MnO on the determination of the N\'{e}el temperature. Moreover, the underlying Heisenberg spin Hamiltonian restricts the system to a fixed lattice. Employing the mHDNNP we can overcome both limitations step by step to reveal their influences on the N\'{e}el temperature. While conventional MC spin-flip simulations use a fixed lattice but allow to explore the specific states of MnO, by including $NpT$ molecular dynamics trajectories in MDMC simulations both the atomic positions as well as the lattice parameters can adapt to the magnetic configurations, and accounting for thermal fluctuations can provide a more realistic description.

To determine the temperature of a phase transition the molar heat capacity can be employed because it shows a peak at the transition temperature for systems of finite size. Employing $NpT$ MD simulations the heat capacity at constant pressure $C_p$ can be obtained from the fluctuations of the total energy, 
\begin{align}
\dfrac{C_p}{k_\mathrm{B}N_\mathrm{A}}=\dfrac{N_\mathrm{atoms}}{k_\mathrm{B}^2T^2N_\mathrm{steps}}\sum_{n=0}^{N_\mathrm{steps}}\Big[E_\mathrm{tot}(n)-\overline{E}_\mathrm{tot}\Big]^2\ ,
\end{align}
with $N_\mathrm{atoms}$ atoms in the simulation cell, the mean simulation temperature $T$, the total energy per atom $E_\mathrm{tot}(n)$ as a function of the MD time step $n$ for the total number of simulation steps $N_\mathrm{steps}$, and the mean total energy per atom during the simulation $\overline{E}_\mathrm{tot}$. $N_\mathrm{A}$ is the Avogadro constant. 

From MC spin-flip simulations the influence of the magnetic degrees of freedom on the heat capacity at constant volume $C_V$ can be calculated,
\begin{align}
\dfrac{C_V}{k_\mathrm{B}N_\mathrm{A}}=3+\dfrac{N_\mathrm{atoms}}{k_\mathrm{B}^2T^2N_\mathrm{steps}}\sum_{n=0}^{N_\mathrm{steps}}\Big[E(n)-\overline{E}\Big]^2\ .
\end{align}
The contribution of the atomic motions to $C_V$ is taken into account by the term $3k_\mathrm{B}N_\mathrm{A}$, as the atomic motions are not considered in the conventional MC spin-flip simulations. The energy fluctuation is calculated from the potential energies per atom $E(n)$ as a function of the MC step $n$ for the total number of steps $N_\mathrm{steps}$ compared to the mean potential energy per atom $\overline{E}$.

To identify the AFM-II to paramagnetic (PM) transition, the temperature dependence of an order parameter $C$ can be used,
\begin{align}
\begin{split}
&C=\dfrac{1}{N_\mathrm{steps}}\sum_{n=0}^{N_\mathrm{steps}}\mathrm{max}\Bigg[\Bigg|\dfrac{\mathbf{s}_\mathrm{(111)}\cdot\mathbf{s}(n)}{|\mathbf{s}_\mathrm{(111)}|\cdot|\mathbf{s}(n)|}\Bigg|,\\
&\left|\dfrac{\mathbf{s}_\mathrm{(\overline{1}11)}\cdot\mathbf{s}(n)}{|\mathbf{s}_\mathrm{(\overline{1}11)}|\cdot|\mathbf{s}(n)|}\right|,\left|\dfrac{\mathbf{s}_\mathrm{(1\overline{1}1)}\cdot\mathbf{s}(n)}{|\mathbf{s}_\mathrm{(1\overline{1}1)}|\cdot|\mathbf{s}(n)|}\right|,\left|\dfrac{\mathbf{s}_\mathrm{(11\overline{1})}\cdot\mathbf{s}(n)}{|\mathbf{s}_\mathrm{(11\overline{1})}|\cdot|\mathbf{s}(n)|}\right|\Bigg]\ ,
\end{split}
\end{align}
with $\mathbf{s}$ being the vector of the spin coordinates of all atoms in the simulation cell. The vector $\mathbf{s}$ at step $n$ is compared with the vectors $\mathbf{s}_\mathrm{(111)}$, $\mathbf{s}_\mathrm{(\overline{1}11)}$, $\mathbf{s}_\mathrm{(1\overline{1}1)}$, and $\mathbf{s}_\mathrm{(11\overline{1})}$ which are the possible AFM-II configurations in different spatial orientations. For normalization both vectors are divided by their lengths. The maximum agreements of the relative magnetic configurations at each step are averaged over the simulation length, i.e., the most similar of the four AFM-II spatial orientations is always used. If the magnetic configuration corresponds to AFM-II during the whole simulation, the order parameter is $C=1$. The order parameter of the second cubic global minimum is $C=0.5$. The variety of paramagnetic configurations leads to a smaller value of the order parameter with $C\lesssim0.2$, since the paramagnetic orders are not correlated to the AFM-II order.

\begin{figure}[htb]
\centering
\includegraphics[width=\columnwidth]{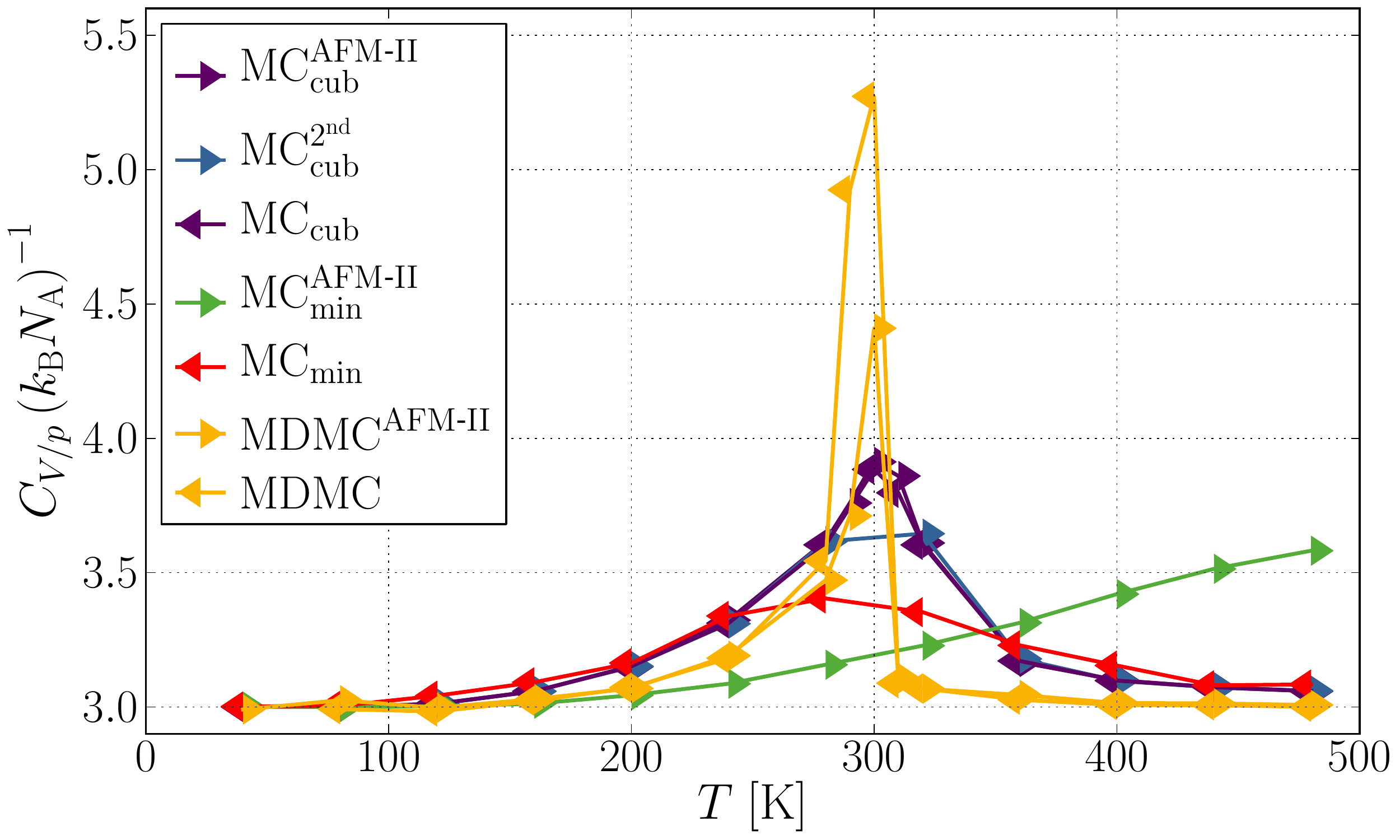}
\caption{Molar heat capacity at constant volume $C_V$ (MC) and pressure $C_p$ (MDMC) as a function of the temperature $T$ obtained in different simulation methods for $6\times6\times6$ MnO supercells. The superscript of the simulation method defines the initial magnetic configuration (cubic AFM-II or $2^\mathrm{nd}$ global minimum), otherwise random initial magnetic orders are used. The subscript cub indicates the restriction to the cubic lattice with $a^\mathrm{exp}=4.430\,\mathrm{\AA}$ and the subscript min indicates an optimization of the initial structure prior to the MC simulation.}\label{fig:heat_capacity}
\end{figure}

\begin{figure}[htb]
\centering
\includegraphics[width=\columnwidth]{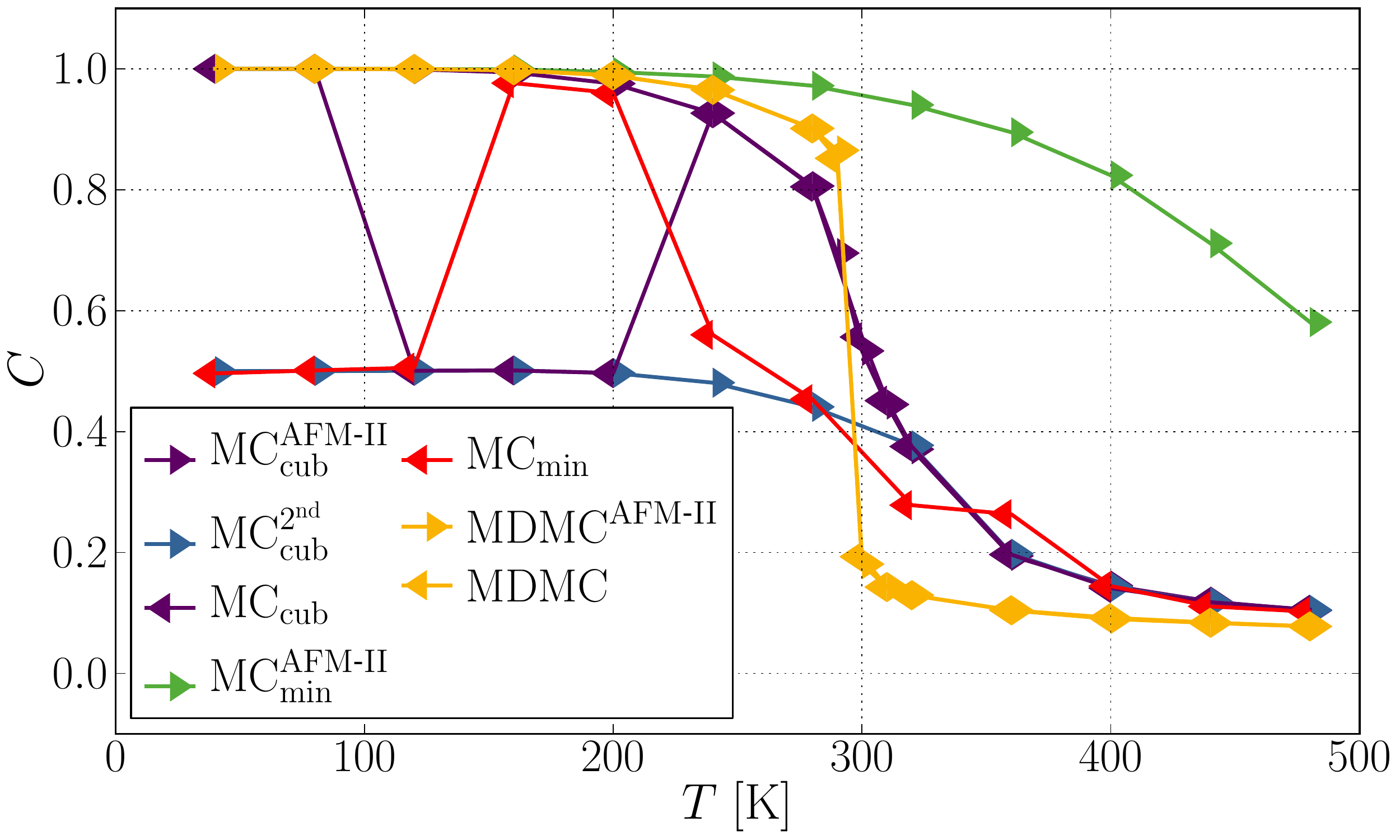}
\caption{Order parameter $C$ as a function of the temperature $T$ obtained in different simulation methods for $6\times6\times6$ MnO supercells. The superscript of the simulation method defines the initial magnetic configuration (cubic AFM-II or $2^\mathrm{nd}$ global minimum), otherwise random initial magnetic orders are used. The subscript cub indicates the restriction to the cubic lattice with $a^\mathrm{exp}=4.430\,\mathrm{\AA}$ and the subscript min indicates an optimization of the initial structure prior to the MC simulation. Lines are only drawn to guide the eyes.}\label{fig:order_parameter}
\end{figure}

MC spin-flip simulations of a cubic $6\times6\times6$ MnO supercell using the experimental lattice constant $a^\mathrm{exp}=4.430\,\mathrm{\AA}$ yield a transition temperature of 300\,K (MC$^{\mathrm{AFM}\text{-}\mathrm{II}}_\mathrm{cub}$ and MC$_\mathrm{cub}$ in Figure \ref{fig:heat_capacity}). The order parameter of these simulations (MC$^{\mathrm{AFM}\text{-}\mathrm{II}}_\mathrm{cub}$ and MC$_\mathrm{cub}$ in Figure \ref{fig:order_parameter}) proves that the transition temperature belongs to the antiferromagnetic to paramagnetic transition. The N\'{e}el temperature is the same for the disordering process (MC$_\mathrm{cub}^{\mathrm{AFM}\text{-}\mathrm{II}}$) as well as the ordering process (MC$_\mathrm{cub}$), i.e., no hysteresis is present. The AFM-II to PM (MC$^{\mathrm{AFM}\text{-}\mathrm{II}}_\mathrm{cub}$) as well as the second cubic global minimum to PM (MC$^{2^\mathrm{nd}}_\mathrm{cub}$) transition temperatures are identical because of the equal interactions of both cubic global minimum configurations. The ordering process can consequently also end in the second cubic global minimum configuration as happened in the MC$_\mathrm{cub}$ simulations at 120, 160, and 200\,K. However, below the N\'{e}el temperature the interconversion of both cubic global minima is kinetically hindered. In summary, the inclusion of specific state information increases the N\'{e}el temperature of MnO by about 50\,K compared to mean field theory.

The restriction to a cubic lattice is an approximation because on the one hand the AFM-II configuration prefers a rhombohedral lattice. Employing this lattice (MC$_\mathrm{min}^{\mathrm{AFM}\text{-}\mathrm{II}}$) increases the N\'{e}el temperature above 480\,K. This overestimation is a consequence of the decreased AFM-II energy and increased PM energy. On the other hand, an optimized lattice of the initial PM configuration (MC$_\mathrm{min}$) leads to a N\'{e}el temperature of about 280\,K. In conclusion, the choice of a specific fixed lattice can change the N\'{e}el temperature at least by 200\,K. Adapting the lattice as well as the atomic positions to the magnetic order is consequently of major importance to obtain reliable results.

\begin{figure}[htb]
\centering
\includegraphics[width=\columnwidth]{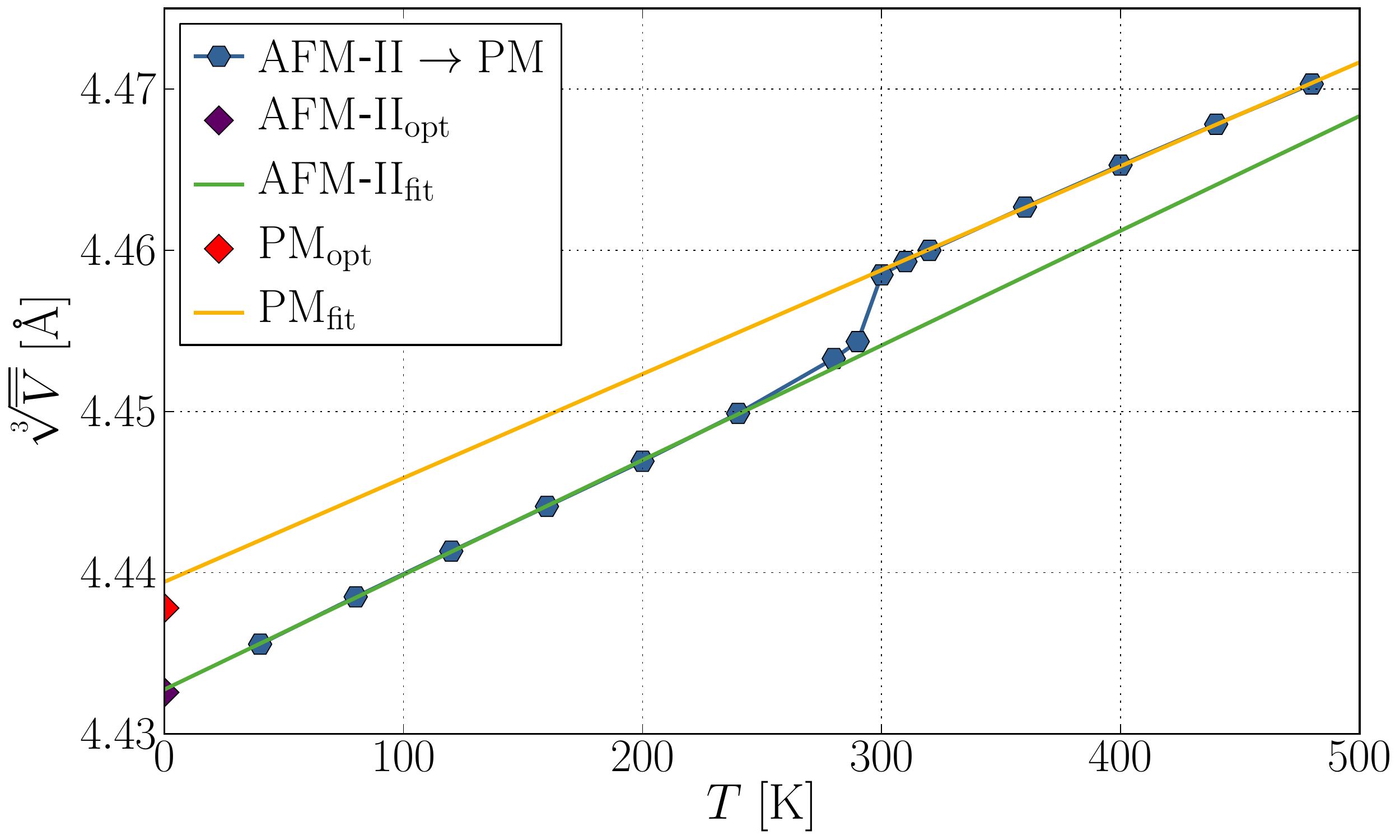}
\caption{Cube root of the mean volume per unit cell $\sqrt[3]{\overline{V}}$ as a function of the temperature $T$ obtained in 10\,ns $NpT$ MD simulations including MC spin-flips of $6\times6\times6$ MnO supercells as well as the optimized values at $p=1\,\mathrm{bar}$ of the AFM-II and PM configuration. The straight lines represent linear fits for the high and low temperature regions.}\label{fig:lattice_volume}
\end{figure}

$NpT$ MD steps enable to sample the thermodynamic equilibrium of the given magnetic configuration at pressure $p$ and temperature $T$ to get rid of simulation artifacts due to a restricted geometric structure. The MDMC simulations are therefore a much better representation of realistic, experimental conditions. MDMC simulations predict a AFM-II to PM transition at 300\,K (see MDMC$^{\mathrm{AFM}\text{-}\mathrm{II}}$ and MDMC in Figure \ref{fig:heat_capacity} and \ref{fig:order_parameter}). This temperature is similar to the result of the cubic lattice due to a compensation of two main influences: The optimized AFM-II lattice (at $p=1\,\mathrm{bar}$) leads to an energy gain of 1.9\,meV\,atom$^{-1}$ compared to the cubic lattice with $a^\mathrm{exp}$ whereby the mean energy gain of the PM configurations is 1.3\,meV\,atom$^{-1}$. The data of the PM phase were calculated from optimization results of 1000 PM configurations, which were obtained during the 10\,ns MDMC simulation at 400\,K. In addition, thermal fluctuations of the atomic positions are included in MDMC simulations leading to thermal expansion of the MnO lattice with increasing temperature. To quantify the energy increase due to the increased lattice volume, optimizations at various pressures of the AFM-II and PM configurations were performed. If the optimized volume at $p=1\,\mathrm{bar}$ is increased to the volume given at 300\,K in the MDMC simulations (see Figure \ref{fig:lattice_volume}), the energy of the AFM-II configuration is increased by 1.7\,meV\,atom$^{-1}$ and the mean energy of the PM configurations by 1.1\,meV\,atom$^{-1}$ (see Supplementary Information \cite{SI}).

The cube root of the mean lattice volume, i.e., a hypothetical averaged cubic lattice constant, shows a discontinuous increase of 0.005\,$\mathrm{\AA}$ at the N\'{e}el temperature (Figure \ref{fig:lattice_volume}). This increase is similar to the difference of the optimized AFM-II and PM values which differ by 0.005\,$\mathrm{\AA}$. Experimentally an increase of about 0.004\,$\mathrm{\AA}$ has been observed at the N\'{e}el temperature \cite{Morosin1970}. The mean lattice angle decreases from 90.77 to 90.00\,$^\circ$ at the N\'{e}el temperature as shown in the Supplementary Information \cite{SI} matching the experimental PM lattice angle of 90.00\,$^\circ$ \cite{Morosin1970}. This discontinuous change of the lattice volume and shape confirms the assignment to a first-order magnetic phase transition \cite{Seino1973, Miyahara1977}.

The linear thermal expansion coefficient of the paramagnetic phase has been measured to be $\alpha_\mathrm{L}^\mathrm{exp}=12\cdot10^{-6}\,\mathrm{K}^{-1}$ at 400\,K \cite{Suzuki1979}. From the MDMC simulations it can be calculated by the change of the lattice constant with temperature at constant pressure divided by the lattice constant,
\begin{align}
\alpha_\mathrm{L}(T)=\dfrac{1}{a(T)}\left(\dfrac{\mathrm{d}a(T)}{\mathrm{d}T}\right)_p\ .
\end{align}
Employing the PM data a value of $\alpha_\mathrm{L}^\mathrm{mHDNNP}=14\cdot10^{-6}\,\mathrm{K}^{-1}$ at 400\,K is obtained in good agreement with experiment.

\subsection*{Defects}

Real materials cannot be considered as infinitely large periodic systems, because they contain surfaces and defects breaking the ideal AFM-II order. For example, Mn vacancies lead to imperfections in the magnetic order and increase the oxidation states of other Mn ions to compensate for the excess of O atoms ensuring overall charge neutrality. The impact of these defects can be predicted by the mHDNNP. In principle, it is also possible to study the role of surfaces as well as doping, but our current parameterization is based on Mn$_x$O bulk data only, with $x=0.969$ and $1$, and thus the present mHDNNP is not applicable to surface structures or doped MnO.

\begin{figure}[htb]
\centering
\includegraphics[width=\columnwidth]{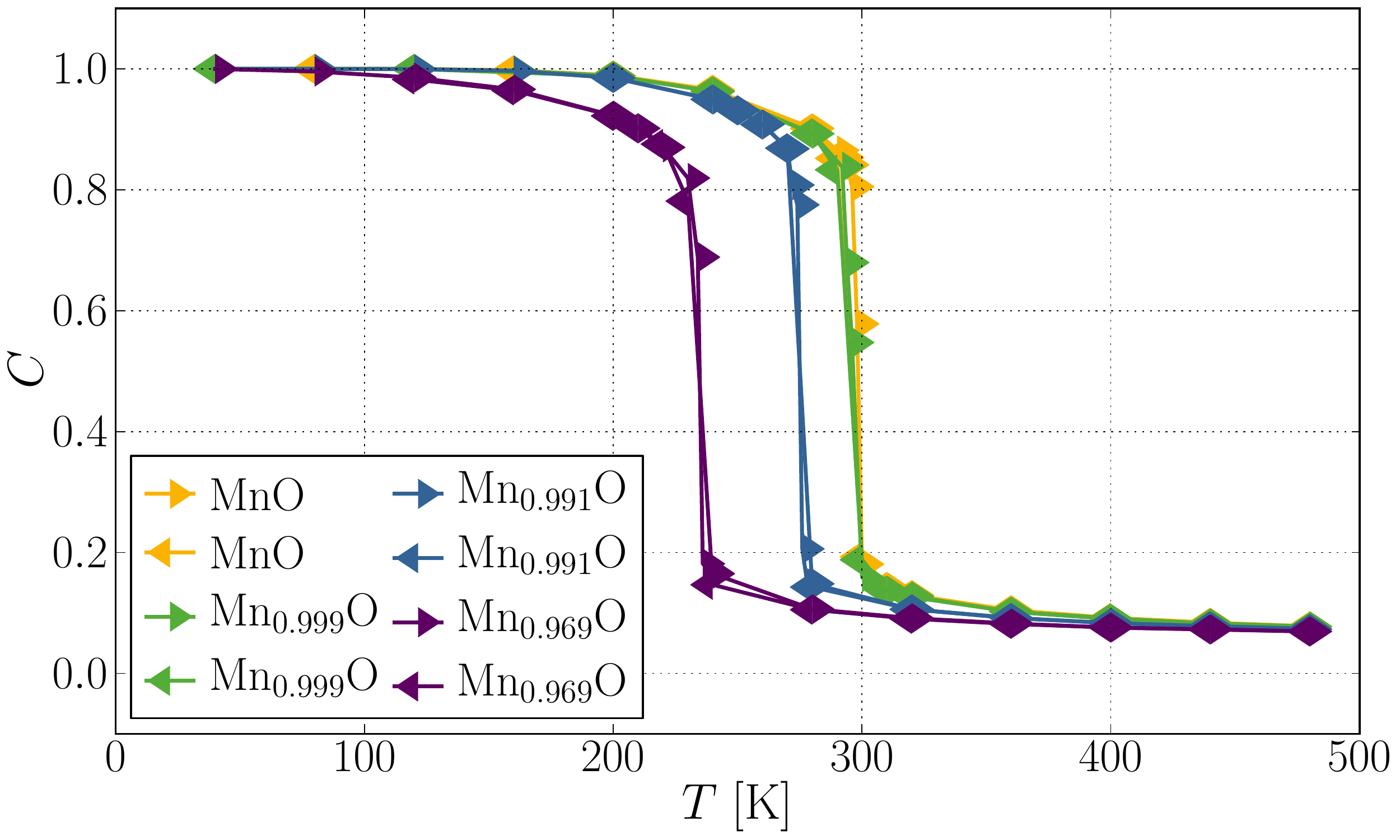}
\caption{Order parameter $C$ as a function of the temperature $T$ obtained in MDMC simulations for $6\times6\times6$ supercells of MnO, Mn$_{0.999}$O, Mn$_{0.991}$O, and Mn$_{0.969}$O.}\label{fig:order_parameter_defect}
\end{figure}

From the heat capacities shown in the Supplementary Information \cite{SI} and the order parameters in Figure \ref{fig:order_parameter_defect} the N\'{e}el temperatures can be determined to be $(298\pm1)\,\mathrm{K}$ for MnO, $(296\pm1)\,\mathrm{K}$ for Mn$_{0.999}$O, $(275\pm1)\,\mathrm{K}$ for Mn$_{0.991}$O, and $(235\pm1)\,\mathrm{K}$ for Mn$_{0.969}$O. Consequently, the increasing Mn vacancy concentration decreases the N\'{e}el temperature, but the change per vacancy is reduced at higher concentrations. The difference between the theoretical and experimental N\'{e}el temperature is therefore not only a consequence of the overestimation by DFT underlying the mHDNNP. A model system containing defects like vacancies, which is more comparable to real, experimental conditions, reduces the difference as well and is therefore required for accurate predictions.

\section*{Conclusion}

In this work we have introduced spin-dependent atom-centered symmetry functions (sACSF), which enable the construction of magnetic high-dimensional neural network potentials (mHDNNP) including the full magnetic and geometric configuration space of spin-polarized, multicomponent systems. Using MnO as model systems we show that energy errors as low as 1\,meV\,atom$^{-1}$ can be reached, which is an order of magnitude lower than employing structure-dependent descriptors only. Structural changes due to the magnetic order are accurately predicted for ground and magnetically excited states with errors of only about 0.001\,$\mathrm{\AA}$ and 0.1\,$^\circ$ compared to the hybrid density functional theory (DFT) reference calculations. Furthermore, the determination of exchange coupling constants, which agree within 0.1\,K with the DFT reference, demonstrates the high quality in the description of the magnetic interactions.

mHDNNPs combine the accuracy and generality of first principles methods with an efficiency close to spin lattice models. The N\'{e}el temperature of MnO calculated by mean field theory differs by about 50\,K from the Monte Carlo result which explicitly samples the magnetic configurations. Including structural fluctuations in the prediction of magnetic transition temperatures is essential because fixing the lattice to the low- or high-temperature configuration can lead to differences of more than 200\,K. mHDNNP-driven molecular dynamics simulations including Monte Carlo spin-flips reveal a small volume increase and the disappearance of the rhombohedral distortion at the N\'{e}el temperature of MnO, as the method is able to provide mean geometric and thermodynamic data of the paramagnetic phase. We find that Mn vacancies lead to a reduced N\'{e}el temperature, which is expected to be relevant for a comparison to experimental results. For instance, changing the stoichiometry from MnO to about Mn$_{0.99}$O the mHDNNP predicts a reduction of the N\'{e}el temperature by about 25\,K. 

Beyond the present work, we expect the mHDNNP method to be a powerful tool for highly accurate, large-scale atomistic simulations of systems involving different spin states, like a variety of magnetic bulk materials, surfaces, and interfaces as well as molecular transition metal complexes containing spin-polarized atoms. Theoretical predictions of the magnetic, geometric, and thermodynamical implications of surfaces, interfaces, defects, and doping can provide interesting control tactics of materials properties and finally for technological applications.

\section*{Methods}

\subsection*{High-Dimensional Neural Network Potential}

The MLP used in this work is a second-generation high-dimensional neural network potential (HDNNP) \cite{Behler2007}. In this method the potential energy $E$ is constructed as a sum of atomic energy contributions,
\begin{align}
E=\sum_{m=1}^{N_{\rm elem}}\sum_{n=1}^{N_{\rm atoms}^{m}}E_n^m(\mathbf{G}^m_n)\ ,
\end{align}
for a system containing $N_{\rm elem}$ elements and $N_{\rm atoms}^{m}$ atoms of element $m$. For each element an individual feed-forward neural network is trained which can provide the atomic energy contribution as a function of the local chemical environment and which is evaluated as often as atoms of the respective element are present in the system. The structural descriptors $\mathbf{G}^m_n$ are vectors of many-body atom-centered symmetry functions (ACSF) \cite{Behler2011}, which fulfill the mandatory translational, rotational, and permutational invariances of the potential energy surface and serve as input vectors of the atomic neural networks. ACSFs describe the local chemical environment of a given central atom as a function of the positions of all neighboring atoms inside a cutoff sphere of radius $R_\mathrm{c}$. To include all energetically relevant interactions the cutoff radius has to be sufficiently large. Besides the positions of the atoms, only the elements have to be specified leading to a reactive potential being able to describe the making and breaking of bonds. The dimensionality of the ACSF vectors can be predefined for each element individually and does not depend on the specific atomic environments. This ensures that the number of input neurons of the atomic feed-forward neural networks remains constant during molecular dynamics simulations. After optimizing the parameters of the neural networks in a training process using the potential energies and atomic force components of reference systems obtained from DFT, the HDNNP can be applied in large-scale simulations at a small fraction of the computational costs. More details about HDNNPs, their construction and validation can be found in several recent reviews \cite{Behler2014, Behler2015, Behler2017, Behler2021}. 

\subsection*{Atom-Centered Symmetry Functions}

Two types of ACSFs are most commonly used for the construction of HDNNPs: The radial symmetry functions
\begin{align}
G_i^\mathrm{rad}=\sum_j\mathrm{e}^{-\eta R_{ij}^2}\cdot f_\mathrm{c}\left(R_{ij}\right)\ ,
\end{align}
and the angular symmetry functions
\begin{align}
\begin{split}
&G_i^\mathrm{ang}=2^{-\zeta}\sum_j\sum_{k\neq j}\left[1+\lambda\cos\left(\theta_{ijk}\right)\right]^\zeta\\
&\cdot\mathrm{e}^{-\eta\left(R_{ij}^2+R_{ik}^2+R_{jk}^2\right)}\cdot f_\mathrm{c}\left(R_{ij}\right)\cdot f_\mathrm{c}\left(R_{ik}\right)\cdot f_\mathrm{c}\left(R_{jk}\right)\ ,
\end{split}
\end{align}
with the cutoff function
\begin{align}
f_\mathrm{c}\left(R_{ij}\right)=\begin{cases}\tfrac{1}{2}\cos\left(\tfrac{\pi R_{ij}}{R_\mathrm{c}}\right)+\tfrac{1}{2}&\mathrm{for}\ R_{ij}\leq R_\mathrm{c}\\0&\mathrm{otherwise}\end{cases}\ .
\end{align}
$R_{ij}$ is the distance between central atom $i$ and neighboring atom $j$, $\theta$ is the angle $j-i-k$ involving two neighbors $j$ and $k$, and $\eta$, $\lambda$, and $\zeta$ are parameters defining the spatial shapes of the ACSFs. Consequently, the ACSF values only depend on the local geometric environment of the atoms. For multicomponent systems containing several elements ACSFs for all element combinations are explicitly included. A detailed discussion of the properties of conventional ACSFs and further functional forms can be found in Reference \cite{Behler2011}.

\subsection*{Spin-Dependent Atom-Centered Symmetry Functions}

To describe the magnetic configuration, we now introduce an atomic spin coordinate
\begin{align}
s_i=\begin{cases}0&\mathrm{for}\ |M_S|<M_S^\mathrm{thres}\\\mathrm{sgn}(M_S)&\mathrm{otherwise}\end{cases}\ ,
\end{align}
with
\begin{align}
M_S=\tfrac{1}{2}(n_\uparrow-n_\downarrow)\ .
\end{align}
$M_S$ is the half-difference of the number of spin-up electrons $n_\uparrow$ and spin-down electrons $n_\downarrow$ of an atom $i$ in a collinear spin-polarized calculation. Consequently, the atomic spin coordinate is equal to the net direction of the atomic spin, i.e., the sign of $M_S$, unless the absolute atomic spin value is smaller than a threshold value $M_S^\mathrm{thres}$, which we introduce to filter out noise in the atomic spin reference data arising from the ambiguity in assigning spins in electronic structure calculations. In this work $M_S^\mathrm{thres}$ is set to 0.25.

The set of atomic spin coordinates can represent all possible collinear magnetic configurations of a system enabling to identify ferro- and antiferromagnetic spin arrangements as well as non-magnetic interactions. Still, we note that different atomic oxidation and spin states with the same spin orientations cannot be distinguished by the spin coordinate alone. Such situations can often be observed, for example, for high- and low-spin states of transition metal ions. However, often the resulting different orbital occupations give rise to structural changes in the local atomic environments like changes in bond lengths or Jahn-Teller distortions and can thus be described by the conventional spatial ACSFs as shown in our previous studies \cite{Eckhoff2020a, Eckhoff2020b} as long as there is a unique relation between the geometric structure and the electronic configuration. In principle, also the inclusion of specific spin values beyond the relative sign might be of interest, but we leave this aspect to future work here.

To integrate the spin coordinates into the radial ACSFs, the radial spin-augmentation function (SAF) $M^\mathrm{x}(s_i,s_j)$ is employed,
\begin{align}
G_i^\mathrm{rad}=\sum_jM^\mathrm{x}(s_i,s_j)\cdot\mathrm{e}^{-\eta R_{ij}^2}\cdot f_\mathrm{c}\left(R_{ij}\right)\ . \label{eq:sACSFrad}
\end{align}
Different radial SAFs, with $\mathrm{x}=0, +, -$, are used to describe the interactions of same (ferromagnetic interactions) and opposite spin directions (antiferromagnetic interactions) respectively,
\begin{align}
M^\mathrm{0}(s_i,s_j)&=1\ ,\\
M^\mathrm{+}(s_i,s_j)&=\tfrac{1}{2}\left|s_is_j\right|\cdot\left|s_i+s_j\right|\ ,\\
M^\mathrm{-}(s_i,s_j)&=\tfrac{1}{2}\left|s_is_j\right|\cdot\left|s_i-s_j\right|\ .
\end{align}

\begin{figure}[htb]
\centering
\includegraphics[width=\columnwidth]{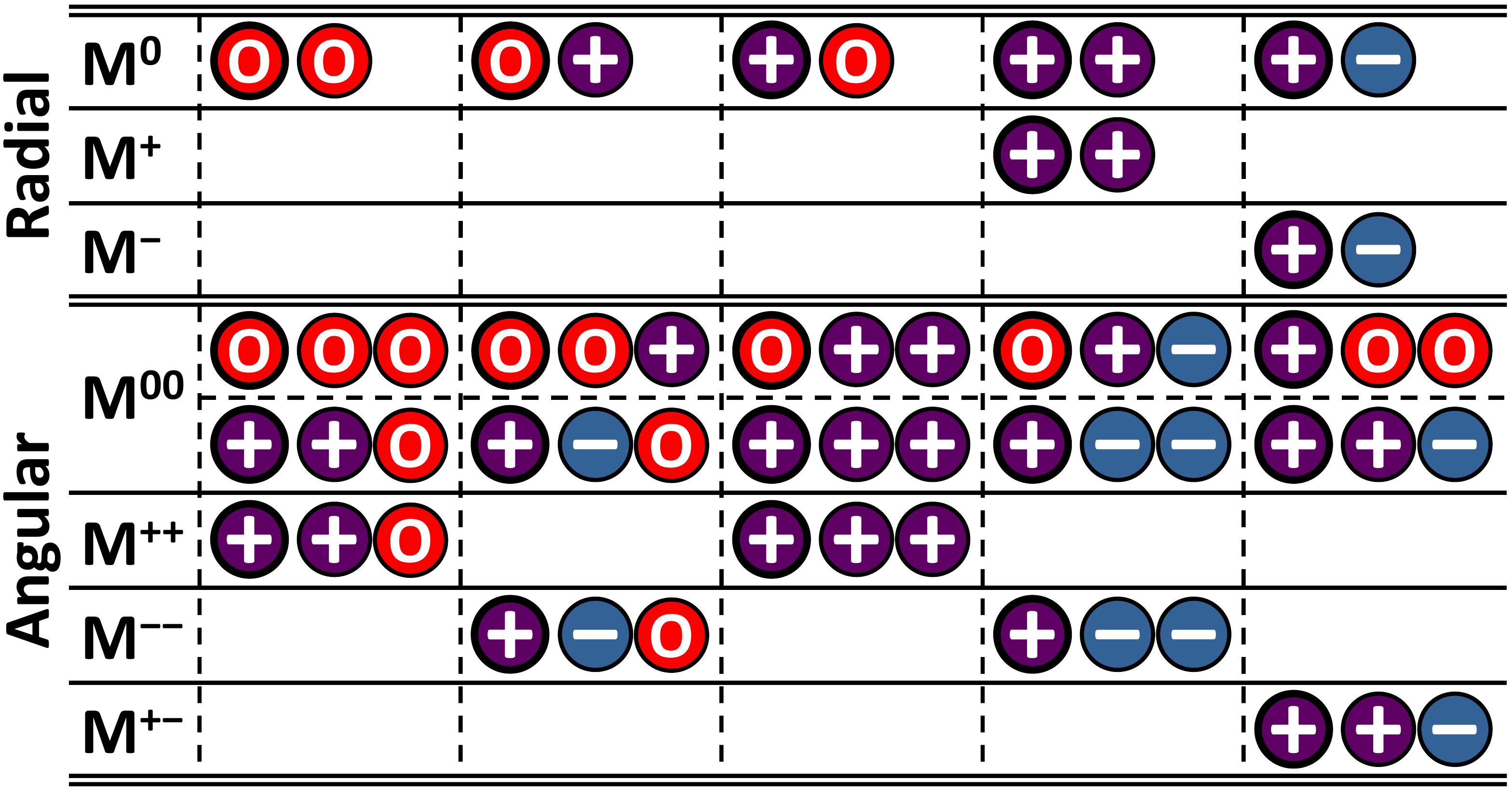}
\caption{Interactions resulting in non-zero values of the radial and angular spin-augmentation functions $M^\mathrm{x}$ and $M^\mathrm{xx}$. Red circles with a zero represent atoms with $s=0$, purple circles with a plus sign atoms with $s=1$, and blue circles with a minus sign atoms with $s=-1$. The first atom of each entry is the central atom $i$ of the sACSF. The order of the neighbor atoms in the angular interactions is insignificant. The inverse interactions (i.e., switching ``+'' and ``-'') yield the same result and are not shown for clarity.}\label{fig:sACSF}
\end{figure}

The radial SAFs $M^\mathrm{x}(s_i,s_j)$ are non-zero only for specific combinations of the spin coordinates of the involved atom pairs as summarized in Figure \ref{fig:sACSF}. This spin augmentation filters the contributions to the radial spin-dependent atom-centered symmetry function (sACSF) in Equation (\ref{eq:sACSFrad}) to distinguish the different magnetic interactions. Only interactions between atoms of the same spin directions (parallel spins) contribute to a sACSF containing $M^\mathrm{+}$ and only interactions between atoms of different spin directions (antiparallel spins) contribute to a sACSF containing $M^\mathrm{-}$. If $s=0$ for one or both of the interacting atoms, $M^\mathrm{+}$ and $M^\mathrm{-}$ are zero leaving only a contribution to the sACSF containing $M^\mathrm{0}$. Taking the absolute values in $M^\mathrm{+}$ and $M^\mathrm{-}$ ensures that the descriptor is only dependent on the relative spin direction. In this way, a simultaneous sign change of all atomic spins does not change the value of the sACSFs ensuring the invariance of the potential energy with respect to the absolute spin orientation. $M^\mathrm{0}$ is used to describe the non-magnetic, purely geometry-dependent interactions between a $s\neq0$ atom and a $s=0$ atom or between two $s=0$ atoms, which are not included in the other terms. 

When constructing a mHDNNP, it is sufficient to use sACSFs only as input for the feed-forward neural networks of elements exhibiting atoms with non-zero spins. For the feed-forward neural networks of all other elements conventional ACSFs can be used. In the same way as ACSFs, sACSFs are constructed for individual element combinations. The choice, which SAFs, i.e., only $M^\mathrm{0}$ or both $M^\mathrm{+}$ and $M^\mathrm{-}$, are required for a given element combination, can be made before constructing the potential because in most systems the atoms of a given element are either all characterized by $s=0$ or by $s\neq0$. For instance, in MnO the manganese atoms exhibit $s\neq0$ and the oxygen atoms correspond to $s=0$. Still, the method is also applicable to other systems including partly magnetically active elements. These systems require the usage of $M^\mathrm{0^*}$ (see Supplementary Information \cite{SI}) instead of $M^\mathrm{0}$ to explicitly separate the non-magnetic interactions from magnetic interactions as well as a careful choice of $M_S^\mathrm{thres}$ to assign physically meaningful spin coordinate values. For element combinations of partly magnetically active elements with (partly) magnetically active elements all radial SAFs $M^\mathrm{0^*}$, $M^\mathrm{+}$, and $M^\mathrm{-}$ are then required.

In a similar way angular sACSFs can be defined as
\begin{align}
\begin{split}
&G_i^\mathrm{ang}=2^{-\zeta}\sum_j\sum_{k\neq j}M^\mathrm{xx}(s_i,s_j,s_k)\cdot\left[1+\lambda\cos\left(\theta_{ijk}\right)\right]^\zeta\\
&\cdot\mathrm{e}^{-\eta\left(R_{ij}^2+R_{ik}^2+R_{jk}^2\right)}\cdot f_\mathrm{c}\left(R_{ij}\right)\cdot f_\mathrm{c}\left(R_{ik}\right)\cdot f_\mathrm{c}\left(R_{jk}\right)\ ,
\end{split}
\end{align}
containing the angular SAFs $M^\mathrm{xx}(s_i,s_j,s_k)$. They allow to distinguish three different interactions of a central $s\neq0$ atom $i$ with two neighboring $s\neq0$ atoms $j$ and $k$: (1) $s_i=s_j=s_k$, (2) $s_i\neq s_j=s_k$, and (3) $s_i=s_j\neq s_k$. The fourth possibility $s_i=s_k\neq s_j$ is equivalent to (3) since the sums over $j$ and $k$ in the angular sACSFs include the interactions $j-i-k$ and $k-i-j$, as both $j$ and $k$ sum over all contributing neighbor atoms to exclude any dependence on the order of the atoms. An efficient separation of these interactions is given by the functions,

\begin{align}
&M^\mathrm{00}(s_i,s_j,s_k)=1\ ,\\
\begin{split}
&M^\mathrm{++}(s_i,s_j,s_k)\\
&=\begin{cases}\tfrac{1}{2}\left|s_i\right|\cdot\left(\left|s_i+s_j+s_k\right|-1\right)&\hspace{-0.15cm}\begin{cases}\mathrm{for}\ s_{j}\neq0\land s_{k}\neq0\\
\mathrm{for}\ s_{j}=0\land s_{k}=0\end{cases}\\
\tfrac{1}{2}\left|s_i\right|\cdot\left|s_i+s_j+s_k\right|&\hspace{-0.15cm}\mathrm{otherwise}\end{cases}\hspace{-0.3cm},
\end{split}\\
\begin{split}
&M^\mathrm{--}(s_i,s_j,s_k)\\
&=\begin{cases}\tfrac{1}{2}\left|s_i\right|\cdot\left(\left|s_i-s_j-s_k\right|-1\right)&\hspace{-0.15cm}\begin{cases}\mathrm{for}\ s_{j}\neq0\land s_{k}\neq0\\
\mathrm{for}\ s_{j}=0\land s_{k}=0\end{cases}\\
\tfrac{1}{2}\left|s_i\right|\cdot\left|s_i-s_j-s_k\right|&\hspace{-0.15cm}\mathrm{otherwise}\end{cases}\hspace{-0.3cm},
\end{split}\\
&M^\mathrm{+-}(s_i,s_j,s_k)=\left|s_is_js_k\right|\cdot\left(\left|s_i+s_j-s_k\right|-1\right)\ ,
\end{align}
as depicted in Figure \ref{fig:sACSF}. 
$M^\mathrm{00}$ is required here to describe interactions including more than one $s=0$ atom or if atom $i$ is $s=0$.

$M^\mathrm{++}$, $M^\mathrm{--}$, and $M^\mathrm{+-}$ yield 1 for the interaction type (1), (2), and (3), respectively, and 0 for the other types in case of $s\neq0$ atoms. For the interactions of two $s\neq0$ atoms with one $s=0$ atom, whereby atom $i$ is $s\neq0$, only $M^\mathrm{++}$ and $M^\mathrm{--}$ are required separating the ferromagnetic and antiferromagnetic interactions similarly as in the radial sACSFs. For systems in which atoms of the same element can be $M_S\neq0$ and $M_S=0$, $M^\mathrm{00^*}$ (see Supplementary Information \cite{SI}) has to be used instead of $M^\mathrm{00}$ to explicitly distinguish non-magnetic and magnetic interactions and the contributions of $M^\mathrm{++}$ and $M^\mathrm{--}$ have to be further split as described in the Supplementary Information \cite{SI}.

Employing sACSFs to calculate the atomic energy contributions of atoms of elements for which at least some atoms in the system are spin-polarized, the mHDNNP is able to distinguish different magnetic configurations and can predict the corresponding potential energies. For elements including exclusively atoms with $M_S=0$ the usual geometry-dependent ACSFs can be applied, which is equivalent to using only $M^0$ and $M^{00}$ in the sACSFs of these elements. Both, the training of only several selected magnetic configurations but also the training of the full magnetic configuration space, are possible in this way.

\subsection*{Computational Details}

The collinear spin-polarized DFT reference calculations were performed employing the Fritz-Haber-Institute ab initio molecular simulations (FHI-aims) code (version 200112.2) \cite{Blum2009, FHIaims}. The screened hybrid exchange-correlation functional HSE06 ($\omega=0.11\,a_0$) \cite{Heyd2003, Heyd2006, Krukau2006} and the ``intermediate'' FHI-aims basis set of numeric atom-centered functions excluding the auxiliary 5g hydrogenic functions were employed. A $\mathbf{\Gamma}$-centered $\mathbf{k}$-point grid of $2\times2\times2$ was applied to calculate the $2\times2\times2$ supercells of MnO (64 atoms without vacancies). The convergence criterion for the self-consistency cycle was set to $10^{-6}$\,eV for the total energies and $10^{-4}$\,eV\,$\mathrm{\AA}^{-1}$ for the forces. Hirshfeld spin moments were used to determine the atomic spins coordinates \cite{Hirshfeld1977}. Further details are given in the Supplementary Information \cite{SI}. An extensive benchmark for manganese oxides employing hybrid DFT functionals can be found in our previous work \cite{Eckhoff2020}.

The sACSFs were implemented in a modified version of the RuNNer code version 1.00 \cite{Behler2015, Behler2017, RuNNer} to construct the mHDNNP. A cutoff radius of $R_\mathrm{c}=10.5\,a_0$ was used. A list of the employed parameters of the $n_G^m$ sACSFs for each element $m$ is given in Tables \ref{tab:radial_sACSFs} and \ref{tab:angular_sACSFs}. The feed-forward neural networks consist of $n_G^m$ input neurons, three hidden layers with 20, 15, and 10 neurons, respectively, and one output neuron. The mHDNNP was trained using the cohesive energies, i.e., the total energy minus the sum of the free atom energies, and using atomic force components obtained from DFT calculations of reference structures in different magnetic states. 90{\%} of these data were used for training, i.e., optimization of the mHDNNP's weight parameters. The remaining data were used as test set. Further details about the training can be found in the Supplementary Information \cite{SI}.

\begin{table}[htb]
\centering
\caption{Employed radial sACSFs with $R_\mathrm{c}=10.5\,a_0$. All combinations of SAFs and symmetry function parameters are used for the given element pairs.}
\begin{tabular}{lllll}
\hline
$\bm{i}$-$j$ && $M^\mathrm{x}$ && $\eta\,/\,a_0^{-2}$ \\
\hline
\textbf{O}-O && $M^0$ && 0, 0.00117, 0.00246, 0.00389, 0.00550 \\
\textbf{O}-Mn && $M^0$ && 0, 0.00369, 0.00882, 0.01636, 0.02809 \\
\hline
\textbf{Mn}-O && $M^0$ && 0, 0.00369, 0.00882, 0.01636, 0.02809 \\
\textbf{Mn}-Mn && $M^+$, $M^-$ && 0, 0.00085, 0.00176, 0.00274, 0.00381 \\
\hline
\end{tabular}
\label{tab:radial_sACSFs}
\end{table}

\begin{table}[htb]
\centering
\caption{Employed angular sACSFs with $\eta=0\,a_0^{-2}$ and $R_\mathrm{c}=10.5\,a_0$. All combinations of SAFs and symmetry function parameters are used for the given element combinations.}
\begin{tabular}{lllllll}
\hline
$\bm{i}$-$j$-$k$ &\,& $M^\mathrm{xx}$ &\,& $\lambda$ &\,\,\,& $\zeta$\\
\hline
\textbf{O}-O-O && $M^\mathrm{00}$ && $-$1, 1 && 1, 2, 4, 16\\
\textbf{O}-O-Mn && $M^\mathrm{00}$ && $-$1, 1 && 1, 2, 4, 16\\
\textbf{O}-Mn-Mn && $M^\mathrm{00}$ && $-$1, 1 && 1, 2, 4, 16\\
\hline
\textbf{Mn}-O-O && $M^\mathrm{00}$ && $-$1, 1 && 1, 2, 4, 16\\
\textbf{Mn}-O-Mn && $M^\mathrm{++}$, $M^\mathrm{--}$ && $-$1, 1 && 1, 2, 4, 16\\
\textbf{Mn}-Mn-Mn && $M^\mathrm{++}$, $M^\mathrm{--}$, $M^\mathrm{+-}$ && $-$1, 1 && 1, 2, 4, 16\\
\hline
\end{tabular}
\label{tab:angular_sACSFs}
\end{table}

mHDNNP-driven molecular dynamics (MD) simulations in combination with Monte Carlo (MC) spin-flips were carried out using the Large-scale Atomic/Molecular Massively Parallel Simulator (LAMMPS) \cite{Plimpton1995, LAMMPS} and the neural network potential package (n2p2) \cite{n2p2} as MD library for potentials generated with RuNNer. The n2p2 code was modified to enable the usage of sACSFs. The MD simulations with MC spin-flips (MDMC) employed $6\times6\times6$ Mn$_x$O supercells referring to the geometric unit cell, with $x=0.969,0.991,0.999,1$, i.e., 1701, 1720, 1727, and 1728 atoms. They were run in the isothermal-isobaric ($NpT$) ensemble at a pressure of $p=1\,\mathrm{bar}$ with a timestep of 1\,fs applying the Nos\'{e}-Hoover thermostat and barostat with coupling constants of 0.1\,ps and 1\,ps, respectively \cite{Nose1984, Hoover1985}. MC spin-flips were performed after each time step. The spin-flip rate is not set to measured or calculated rates to study dynamic properties but to sample the thermodynamic equilibrium efficiently. In all MDMC simulations the system was equilibrated for 1\,ns before the acquisition period of 10\,ns. In all conventional MC spin-flip simulations, i.e., no MD steps in between the MC spin-flips, the equilibration was performed for $10^6$ steps and the acquisition consisted of $10^7$ steps.

\section*{Data Availability}

The datasets generated and analyzed during the current study are available from the corresponding author on reasonable request.

\section*{Code Availability}

The modified versions of RuNNer and n2p2 to enable the usage of sACSFs are available from the corresponding author on reasonable request. The modifications will be implemented in coming release versions under the GPL3 license.

\section*{Acknowledgments}

This project was funded by the Deutsche Forschungsgemeinschaft (DFG) - project No. 217133147/SFB 1073, project C03. We gratefully acknowledge computing time provided by the Paderborn Center for Parallel Computing (PC$^2$) and by the DFG (INST186/1294-1 FUGG, project No.\ 405832858).

\section*{Author Contributions}

M.E.\ conceived the sACSF approach and initiated the project. M.E.\ worked out and implemented the practical algorithms and performed all calculations. M.E.\ and J.B.\ contributed ideas to the project and analyzed the results. M.E.\ wrote the initial version of the manuscript and prepared the figures. M.E.\ and J.B.\ jointly edited the manuscript.\\\\

\section*{Competing Interests}

The authors declare no competing interests.

\bibliographystyle{naturemag}
\bibliography{bibliography}

\end{document}